\documentclass[onecolumn,floatfix,aps,nofootinbib]{revtex4}
\pdfoutput=1
\usepackage[dvips]{epsfig}
\usepackage[english]{babel}
\usepackage[utf8]{inputenc}
\usepackage[T1]{fontenc}
\usepackage{lipsum}       
\usepackage{bbm}
\usepackage{verbatim}
\usepackage{array}
\usepackage{bm} 
\usepackage{amsmath}
\usepackage{yfonts}
\usepackage{amsthm}
\usepackage{amsmath,amscd}
\usepackage{pst-plot}
\usepackage{slashed} 
\usepackage{tikz-cd}
\usepackage{amsfonts}
\usepackage{graphicx}
\usepackage{amssymb}
\newtheorem{definition}{Definition} 
\usepackage{cases}
\usepackage{lmodern}
\usepackage[T1]{fontenc}
\usepackage{textcomp}

\usepackage{indentfirst} 
\usepackage[titletoc]{appendix}
\usepackage{subeqnarray}
\usepackage{indentfirst} 

\usepackage{xcolor}
\usepackage{calrsfs}

\usepackage{wasysym}
\usepackage[all]{xy}
\usepackage{tikz-cd}

    \usepackage{amsmath,amsfonts,amssymb,amsthm,mathrsfs,bbm,braket}

    \numberwithin{equation}{section}

\usepackage{bbold}

\usepackage{hyperref}
\usepackage{setspace}
\begin{document}

\title{From Lorentz to $SIM(2)$: contraction, four-dimensional algebraic relations and projective representations}  

\author{J. E. Rodrigues} 
\email{jose.er.batista@unesp.br}
\affiliation{Departamento de F\'isica, Universidade
	Estadual Paulista, UNESP, Av. Dr. Ariberto Pereira da Cunha, 333, Guaratinguet\'a, SP,
	Brazil.}

\author{J. M. B. Matzenbacher} 
\email{j.matzenbacher@unesp.br}
\affiliation{Departamento de F\'isica, Universidade
	Estadual Paulista, UNESP, Av. Dr. Ariberto Pereira da Cunha, 333, Guaratinguet\'a, SP,
	Brazil.}

\author{G. M. Caires da Rocha} 
\email{gabriel.marcondes@unesp.br}
\affiliation{Departamento de F\'isica, Universidade
	Estadual Paulista, UNESP, Av. Dr. Ariberto Pereira da Cunha, 333, Guaratinguet\'a, SP,
	Brazil.}

\author{J. M. Hoff da Silva} 
\email{julio.hoff@unesp.br}
\affiliation{Departamento de F\'isica, Universidade
Estadual Paulista, UNESP, Av. Dr. Ariberto Pereira da Cunha, 333, Guaratinguet\'a, SP,
Brazil.}

\begin{abstract}
We present a comprehensive study on $SIM(2)$ and $ISIM(2)$ groups, their representations and algebraic aspects. These groups, together with $HOM(2)$, arise as the symmetry groups of Very Special Relativity (VSR), where full Lorentz invariance is reduced while retaining many relativistic consequences. After obtaining $SIM(2)$ through the Inönü-Wigner contraction procedure, a complete four-dimensional algebraic representation is shown for $\mathfrak{sim(2)}$ and $\mathfrak{isim(2)}$. Besides that, we apply Bargmann's formalism to investigate the (projective) representations for both cases, keeping track of the source of phase factors. We complete the study by presenting a particularly simple analysis to probe the existence of local phase factors, which is useful when dealing with non-abelian groups.    
\end{abstract}		

\maketitle

\tableofcontents

\newpage     
\section{Introduction}

Identifying projective representations for symmetry groups has been a subject of interest since the very fundamental work of Wigner \cite{wig1}. The analysis has gained the status of a theory after the comprehensive work of Bargmann \cite{barg}. Understandably, attention was devoted to Galilei, Lorentz, and Poincar\'e groups, but a plethora of subgroups have been known since the middle of the seventies \cite{patera}. Among these subgroups, two Lorentz subgroups, $SIM(2)$ and $HOM(2)$\footnote{The groups $\mathrm{SIM}(2)$ and $\mathrm{HOM}(2)$ stand for similitude and homothety, respectively. These groups receive their names due to the existence of an isomorphism between their Lie algebras and the Lie algebras of similitudes and homotheties of the Euclidean plane.}, had appeared as potentially relevant since their lower symmetries were shown to be enough to explain the negative Michelson-Morley experiment result \cite{cg}. These groups entail a privileged direction in spacetime, and the relativistic theory constructed upon their symmetries is the so-called Very Special Relativity (VSR). 

Regarding low energy effects coming from (VSR), it was shown that $HOM(2)$ and $SIM(2)$ lead to an unobserved magnitude concerning Thomas precession \cite{das,gan}. High energy prospects were addressed in the literature (although not exhaustively) from supersymmetric effects \cite{susy1,susy2} to gauge theories impacts \cite{ger,su,nayak,bufalo}. The impact of relativistic symmetries endowed with a preferred direction in how a particle is described and its quantum field counterpart is constructed was investigated in \cite{cyl}. Furthermore, recent works has established explicit links between deformations of VSR and Finsler geometry \cite{disimb2_finsler,Dimakis,lv_finsler}, formulated $SIM(2)$-covariant renormalization and infrared-regularization frameworks for VSR gauge theories and QED \cite{renorm_vsr_gauge,vsr_ir_loops}, and computed radiative corrections with potentially anisotropic imprints that are stringently constrained by cosmic-photon observations \cite{vsr_vacpol,vsr_lightlight,cosmic_photons}.

The isotropy group associated with $SIM(2)$ is $SO(2)$ for massless and massive particles, a characteristic providing relevant aspects for the construction of irreducible representation as well as an additional richness for mathematical inclined analysis. The $\mathfrak{sim(2)}$ algebra is composed by four generators, $T_1$, $T_2$, $J_3$, and $K_3$, where $J_i$ stands for rotation and $K_i$ for boosts generators ($i=1,2,3$), $T_1=K_1+J_2$, and $T_2=K_2-J_1$. These generators obey the following algebra:
\begin{equation}
[T_1, T_2] = 0; \ \ \ \ \ \ [T_1, K_3] = iT_1; \ \ \ \ \ \ [T_1, J_3] = -iT_2; \nonumber
\end{equation}
\begin{equation}
[T_2, K_3] = iT_2; \ \ \ \ \ \ [T_2, J_3] = iT_1; \ \ \ \ \ \ [K_3, J_3] = 0. \nonumber
\end{equation}

Spacetime translation may be encompassed into the group structure of $SIM(2)$ through semi-direct product resulting in its inhomogeneous counterpart, frequently denoted by $ISIM(2)$. We focus on presenting a comprehensive study about $SIM(2)$ group containing its emergence via a specific Inönü-Wigner contraction, a detailed achievement of $\mathfrak{sim(2)}$ and $\mathfrak{isim(2)}$ four-dimensional algebra, and a complete discussion on the projective character of their representations. Let us briefly contextualize every such achievement: Ultimately, Lie group contractions are potent tools in theoretical and mathematical physics, unveiling how diverse symmetry groups emerge as appropriate limits of more general ones. The example of transitioning from Poincaré to Galilean symmetry remains a canonical demonstration. However, similar contractions underpin numerous contexts in the study of relativistic and non-relativistic field theories, high-energy physics, and beyond. It is indeed insightful to present $SIM(2)$ as a Lorentz contraction result, not just from the strict academic point of view, but it can also be relevant in gauging spacetime symmetries. On the other hand, to our knowledge, there is a lack of literature addressing the construction of a four-dimensional matrix of generators for the $\mathfrak{(i)sim(2)}$ algebra, similar to the approach used for the representations of the Lorentz and Poincaré generator's algebra. We believe its absence has obstructed a more straightforward approach on several grounds, including (but not restricted to) a secure program of gauging its symmetries.    

Finally, a deep understanding of the type of representation for a given group is of extreme relevance in mathematical physics. Although modern techniques from algebraic topology provide an exhaustive tool, more often than never, it is important to apply a more conservative approach to keep track of the origin of the phase factor, leading to a local projective representation. Taking advantage of the founded algebraic closed four-dimensional form, we apply Bargmann's original theory (contrasting it with the central extensions) to pinpoint the source of the local phase factor in $SIM(2)$ representations. Once this aspect is unveiled, we apply standard cohomology analysis to extend the results to $ISIM(2)$. We also discuss a direct method to envisage the presence of local phase factors for non-abelian groups. 

This work is organized as follows: Section II starts discussing general aspects of the Inönü-Wigner contraction, after which the specific $SIM(2)$ case is approached. Section III is devoted to presenting the four-dimensional closed representations to $\mathfrak{sim(2)}$ and $\mathfrak{isim(2)}$. In Section IV, after reviewing the fundamental aspects of Bargmann's theory, we delve into the group representation for $SIM(2)$ and $ISIM(2)$, addressing all their concerns. Also, in Section IV, we present a complementary point of view allowing for a rapid inspection of the existence of local phase factors. In Section V we conclude.    

We do not always notationally discriminate between Lie groups and corresponding Lie algebras, but such a difference is evinced in the context whenever necessary. Moreover, the $\mathfrak{g}$ algebra of a Lie group $G$ is linearly mapped into linear combinations of (infinitesimal) representation generators of $G$, such that $[c_1,c_2]\mapsto [G_{c_1},G_{c_2}]$. This mapping is implied throughout this paper. 

\section{Inönü-Wigner contraction}

The concept of Lie group contraction, introduced by E. Inönü and E. Wigner in 1953 \cite{iw}, provides a systematic procedure to derive a non-isomorphic Lie group from a given one via a well-defined limiting operation. Their seminal contribution demonstrated how the inhomogeneous Galilei group emerges from the Poincaré group in the non-relativistic limit, \(\,c \to \infty\), where $c$ stands for the velocity of light. This remarkable transition elucidates how certain fundamental symmetries underlying relativistic theories reduce to those characterizing classical mechanics under appropriate limits.

\subsection{Procedure Description}

To set the stage, recall that a Lie group \(G\) is naturally endowed with an \(n\)-dimensional differentiable manifold structure. For each point \(g \in G\), one can associate a tangent space \(T_g G\), which is a vector space encompassing all tangent vectors at \(g\). A vector field on \(G\) assigns to every point \(g\) a corresponding vector in \(T_g G\). In particular, let \(\{X_i\}\) with \(i=1,2,\dots,n\) denote a basis for \(T_g G\). Since \(G\) is a Lie group, the elements \(\{X_i\}\) constitute a base for the Lie algebra associated with \(G\). Any group element of \(G\) can then be locally expressed as 
\begin{equation}
{\bf v} \;=\; a^i \, X_i,
\end{equation}
where \(a^i\) are the \(n\) group transformation parameters. These generators \(X_i\) satisfy the Lie bracket's defining relations
\begin{equation}
\label{L.B.}
[X_i, X_j] \;=\; C_{ij}^{\phantom{aa}k}\, X_k,
\end{equation}
with \(C_{ij}^{\phantom{aa}k}\) denoting the structure constants of the algebra. One may perform a non-singular linear transformation \(\,\mathbb{U}\) on the generators \(\{X_i\}\), thus obtaining a new set of generators \(\{Y_i\}\)
\begin{equation}
\label{geradoresY}
Y_i \;=\; \mathbb{U}_i^{\phantom{.}j}\, X_j,
\end{equation}
along with the corresponding redefinition of the group parameters
\begin{equation}
b^i \;=\; a^j\, (\mathbb{U}^{-1})_j^{\phantom{.}i}.
\end{equation}
In order for the new basis \(\{Y_i\}\) to describe the same Lie algebra, the structure constants must transform according to
\begin{equation}
\mathbb{C}_{ij}^{\phantom{aa}k} 
\;=\; \mathbb{U}_i^{\phantom{.}a} \,\mathbb{U}_j^{\phantom{.}b} \, C_{ab}^{\phantom{aa}c}\, (\mathbb{U}^{-1})_c^{\phantom{.}k}.
\end{equation}
Hence, one can see that the same Lie algebra can be described with different sets of generators and those corresponding to different structure constants. As long as \(\mathbb{U}\) is non-singular, \(\{Y_i\}\) remains an equivalent characterization of the same Lie group \(G\). However, if the matrix \(\mathbb{U}\) degenerates (i.e., becomes singular) in a certain limit, the resulting limiting algebra may correspond to a no longer isomorphic Lie group. This fact, combined with Inönü and Wigner's claim --- ``\textit{Every Lie group can be contracted with respect to any of its continuous subgroups, and only with respect to these}'' --- encapsulates the proposed contraction procedure essence.

To explicitly implement the contraction method, one typically allows \(\mathbb{U}\) to depend on a small real parameter \(\epsilon\). A representative scenario treats \(\mathbb{U}\) as a linear function of \(\epsilon\):
\begin{equation}
\label{matrizU}
\mathbb{U}_i^{\phantom{.}j} \;=\; u_i^{\phantom{.}j} \;+\; \epsilon\, w_i^{\phantom{.}j},
\end{equation}
where \(u_i^{\phantom{.}j}\) and \(w_i^{\phantom{.}j}\) are chosen such that \(\mathbb{U}\) remains invertible for all \(\epsilon \neq 0\), but its determinant vanishes as \(\epsilon \to 0\). Under suitable conditions \cite{iw}, the \(\mathbb{U}\)-Matrix may be simplified to its so-called ``normal form'' which decomposes it into block matrices
\begin{equation}
\label{explicitmatrixforms}
u 
\;=\; 
\begin{pmatrix}
\mathbb{I}_{r\times r} & 0_{r\times(n-r)} \\
0_{(n-r)\times r} & 0_{(n-r)\times (n-r)}
\end{pmatrix}, 
\quad
w 
\;=\; 
\begin{pmatrix}
v_{r\times r} & 0_{r\times(n-r)} \\
0_{(n-r)\times r} & \mathbb{I}_{(n-r)\times (n-r)}
\end{pmatrix},
\end{equation}
where \(0 < r < n\), \(\mathbb{I}\) and 
\(0\) corresponds to the identity matrix and the null matrix of appropriate dimensions, respectively. Besides, \(v_{r\times r}\) is an arbitrary \(r\times r\) matrix.

Applying the normal-form expression of \(\mathbb{U}\) to the original generators \(\{X_i\}\) via Eq. \eqref{geradoresY}, we have 
\begin{align}
\label{Y1}
Y_{1\mu} 
&= X_{1\mu} \;+\; \epsilon \,\sum_{\nu=1}^r v_\mu^{\phantom{a}\nu}\,X_{1\nu} 
\qquad (\mu,\nu = 1,2,\dots,r), \\
\label{Y2}
Y_{2\lambda} 
&= \epsilon\, X_{2\lambda} 
\qquad (\lambda = r+1, r+2,\dots,n).
\end{align}
Here, the subscripts ``1'' and ``2'' stand for blocks of the original generator basis \(\{X_i\}\), where `1' refers to the first \(r\) elements, and the subscript `2' to the last \(n-r\) elements. The behavior of these expressions as \(\epsilon \to 0\) precisely indicates how the limiting Lie algebra structure may develop. To examine the closure of the new generator basis \(\{Y_i\}\) under commutation relations, one first analyze the bracket \([Y_{1\mu}, Y_{1\nu}]\). The extension to other Lie Bracket combinations, such as \([Y_{1\mu}, Y_{2\nu}]\) and \([Y_{2\mu}, Y_{2\nu}]\), follows immediately. Beginning with \([Y_{1\mu}, Y_{1\nu}]\), substituting Eq. \eqref{Y1} and retaining terms up to \(\mathcal{O}(\epsilon)\) yields
\begin{equation}
\label{comuY1Y1}
[Y_{1\mu}, Y_{1\nu}] 
\;=\; [X_{1\mu}, X_{1\nu}] 
\;+\; \epsilon \,v_\nu^{\phantom{a}\alpha}\,[X_{1\mu}, X_{1\alpha}] 
\;+\; \epsilon \,v_\mu^{\phantom{a}\alpha}\,[X_{1\alpha}, X_{1\nu}] 
\;+\; \mathcal{O}(\epsilon^2).
\end{equation}
Similarly, one can derive the inverse relations from the invertibility property of \(\mathbb{U}\)
\begin{align}
\label{X1}
X_{1\mu} 
&= Y_{1\mu} \;-\; \epsilon \,v_\mu^{\phantom{a}\nu}\,Y_{1\nu}, 
\qquad (\mu,\nu=1,2,\dots,r), \\
\label{X2}
X_{2\lambda} 
&= \frac{1}{\epsilon}\,Y_{2\lambda}, 
\qquad (\lambda=r+1,r+2,\dots,n).
\end{align}
Now, using equations \eqref{X1}--\eqref{X2}, recalling the original Lie Bracket \eqref{L.B.} for \(\{X_i\}\), one obtains
\begin{eqnarray}
\label{comY1Y1igualY}
[Y_{1\mu}, Y_{1\nu}] 
\;&=&\left.\; C_{1\mu, 1\nu}^{\phantom{aaaa}1\alpha} \,Y_{1\alpha}
\;+\; \frac{1}{\epsilon}\,C_{1\mu, 1\nu}^{\phantom{aaaa}2\alpha}\,Y_{2\alpha}
\;+\; v_{\nu}^{\phantom{a}\beta}\,C_{1\mu, 1\beta}^{\phantom{aaaa}2\alpha}\,Y_{2\alpha}
\right.\nonumber\\
\;&+&\left.\; v_{\mu}^{\phantom{a}\beta}\,C_{1\beta, 1\nu}^{\phantom{aaaa}2\alpha}\,Y_{2\alpha}
\;+\; \mathcal{O}(\epsilon),\right.
\end{eqnarray}
which only converges in the \(\epsilon \to 0\) regime if the condition 
\begin{equation}
\label{convergencecondition}
C_{1\mu, 1\nu}^{\phantom{aaaa}2\alpha} \;=\; 0
\end{equation}
holds, ensuring the coefficient of \(1/\epsilon\) vanishes. Recalling that, if \(\mathbb{U}\) is singular, its image forms a new Lie group, and since it may be expanded in the \(\{Y_i\}\) basis
\[
[Y_{1\mu}, Y_{1\nu}]
\;=\; \mathbb{C}_{1\mu, 1\nu}^{\phantom{aaaa}1\lambda}\,Y_{1\lambda} 
\;+\; \mathbb{C}_{1\mu, 1\nu}^{\phantom{aaaa}2\lambda}\,Y_{2\lambda}.
\]
Comparing this last expression with Eq. \eqref{comY1Y1igualY} in the limit \(\epsilon \to 0\) yields
\begin{equation}
\label{resultado1}
\mathbb{C}_{1\mu, 1\nu}^{\phantom{aaaa}1\lambda} \;=\; C_{1\mu, 1\nu}^{\phantom{aaaa}1\lambda},
\quad\quad 
\mathbb{C}_{1\mu, 1\nu}^{\phantom{aaaa}2\lambda} \;=\; 0 
\;=\; C_{1\mu, 1\nu}^{\phantom{aaaa}2\lambda}.
\end{equation}
Analogous calculations for \([Y_{1\mu}, Y_{2\nu}]\) and \([Y_{2\mu}, Y_{2\nu}]\) show
\begin{align}
[Y_{1\mu}, Y_{2\nu}] 
&= C_{1\mu, 2\nu}^{\phantom{aaaa}2\lambda}\,Y_{2\lambda} 
\;+\; \epsilon\,C_{1\mu, 1\nu}^{\phantom{aaaa}1\lambda}\,Y_{1\lambda}  
\;+\; \mathcal{O}(\epsilon), \\
[Y_{2\mu}, Y_{2\nu}] 
&= \mathcal{O}(\epsilon^2),
\end{align}
and hence
\begin{equation}
\label{resultado2}
\mathbb{C}_{1\mu,2\nu}^{\phantom{aaaa}2\lambda} \;=\; C_{1\mu,2\nu}^{\phantom{aaaa}2\lambda},
\quad\quad
\mathbb{C}_{1\mu, 2\nu}^{\phantom{aaaa}1\lambda} \;=\; 0,
\end{equation}
as well as
\begin{equation}
\label{resultado3}
\mathbb{C}_{2\mu, 2\nu}^{\phantom{aaaa}1\lambda} \;=\; 0,
\quad\quad 
\mathbb{C}_{2\mu, 2\nu}^{\phantom{aaaa}2\lambda} \;=\; 0.
\end{equation}

Collectively, equations \eqref{resultado1} -- \eqref{resultado3} bring together the main aspects of the contraction method procedure: from the original Lie group \(G\) with structure constants \(C_{ij}^{\phantom{..}k}\), one obtains a new, non-isomorphic Lie group, \(G'\), spanned by \(\{Y_i\}\) and defined by the structure constants \(\mathbb{C}_{ij}^{\phantom{..}k}\). The first set of identities, \eqref{resultado1}, reveals that \(G'\) contains a subgroup isomorphic to a corresponding subgroup of \(G\). Conditions \eqref{resultado3} indicate that the generators \(\{X_{2\mu}\}\) defines an abelian subalgebra at the given limit. Finally, the pair \eqref{resultado2} confirms that this abelian subalgebra is invariant under commutations with the subgroup generated by \(\{X_{1\mu}\}\). These properties highlight how contraction manipulates the original symmetry content to yield a distinctly new but systematically related symmetry structure.

\subsection{\texorpdfstring{$SIM(2)$}{} as a contraction of Lorentz Group}
\label{sec:LorentzSIM2}

Now, we outline a method for contracting the Lorentz group with respect to the $SIM(2)$ subgroup, thereby obtaining a lower-dimensional non-isomorphic group in the limit of a singular transformation. In addition, we provide a broader interpretation of how the contraction arises from a suitable choice of base transformation. Although finding an explicit \(\epsilon\)-dependent matrix \(\mathbb{U}\) connecting the Lorentz and $SIM(2)$ bases can be nontrivial, the existence of such a matrix follows directly from the Inönü-Wigner theorem since $SIM(2)$ is indeed a continuous subgroup of the Lorentz group.


We begin by considering the standard Lorentz algebra generators, arranged in the conventional order
\[
X'_i \;=\; (\,J_1,\; J_2,\; J_3,\; K_1,\; K_2,\; K_3\,)^\mathrm{T},
\quad
i \;=\; 1,2,\dots,6.
\]
Here, \(J_i\) (\(i=1,2,3\)) denotes the infinitesimal generators of spatial rotations, while \(K_i\) (\(i=1,2,3\)) correspond to the generators of boosts. By definition, $SIM(2)$ is a Lorentz subgroup; thus, as per the Inönü-Wigner theorem, it satisfies the necessary condition for contraction. Concretely, there exists an \(\epsilon\)-dependent matrix \(\mathbb{U}(\epsilon)\) such that in the limit \(\epsilon \to 0\), one obtains a non-isomorphic Lie algebra embedding the $\mathfrak{sim}(2)$ algebra.

As remarked before, a key step in simplifying the contraction is to perform a preliminary (non-singular) linear transformation on the standard Lorentz basis, which brings the generators into a more amenable form to block-diagonalization and subsequent contraction. Specifically, define a new set of generators
\[
\begin{pmatrix}
T_1\\[6pt] T_2\\[6pt] J_3\\[6pt] K_3\\[6pt] \widetilde{T}_1\\[6pt] \widetilde{T}_2
\end{pmatrix}
\;=\;
\underbrace{
	\begin{pmatrix}
	0 & 1 & 0 & 1 & 0 & 0\\
	-1 & 0 & 0 & 0 & 1 & 0\\
	0 & 0 & 1 & 0 & 0 & 0\\
	0 & 0 & 0 & 0 & 0 & 1\\
	0 & -1 & 0 & 1 & 0 & 0\\
	1 & 0 & 0 & 0 & 1 & 0
	\end{pmatrix}
}_{\displaystyle \mathcal{A}}
\begin{pmatrix}
J_1\\[6pt] J_2\\[6pt] J_3\\[6pt] K_1\\[6pt] K_2\\[6pt] K_3 
\end{pmatrix}.
\label{mat}\]
The matrix \(\mathcal{A}\) is invertible and thus describes an automorphism of the Lorentz algebra. Hence, the set
\[
\{\,T_1,\; T_2,\; J_3,\; K_3,\; \widetilde{T}_1,\; \widetilde{T}_2\},
\]
forms an isomorphic representation of the Lorentz algebra. Concretely, the newly introduced generators are
\[
T_1 \;=\; K_1 \;+\; J_2,
\quad
T_2 \;=\; K_2 \;-\; J_1,
\quad
\widetilde{T}_1 \;=\; K_1 \;-\; J_2,
\quad
\widetilde{T}_2 \;=\; K_2 \;+\; J_1.
\]
All commutation relations in this new basis can be derived from those of the Lorentz algebra. A convenient summary is provided in Table \ref{autoLoralgebratable}.

\begin{table}[h!]
	\centering
	\begin{tabular}{|cr|cr|cr|}
		\hline
		\(\,[T_1,\, T_2]\) \;=\;& \(0\) 
		& \(\,[T_1,\,K_3]\) \;=\;& \(i\,T_1\) 
		& \(\,[\widetilde{T}_1,\,K_3]\) \;=\;& \(i\,\widetilde{T}_1\) 
		\\[6pt]
		\(\,[\widetilde{T}_1,\, \widetilde{T}_2]\) \;=\;& \(0\) 
		& \(\,[T_2,\,K_3]\) \;=\;& \(i\,T_2\) 
		& \(\,[\widetilde{T}_2,\,K_3]\) \;=\;& \(i\,\widetilde{T}_2\)
		\\[6pt]
		\(\,[T_1,\, \widetilde{T}_1]\) \;=\;& \(-\,2\,i\,K_3\)
		& \(\,[T_1,\,J_3\phantom{.}]\) \;=\;& \(-\,i\,T_2\)
		& \(\,[\widetilde{T}_1,\,J_3\phantom{.}]\) \;=\;& \(-\,i\,\widetilde{T}_2\)
		\\[6pt]
		\(\,[T_1,\, \widetilde{T}_2]\) \;=\;& \(-\,2\,i\,J_3\) 
		& \(\,[T_2,\,J_3\phantom{.}]\) \;=\;& \(i\,T_1\)  
		& \(\,[\widetilde{T}_2,\,J_3\phantom{.}]\) \;=\;& \(i\,\widetilde{T}_1\)
		\\[6pt]
		\(\,[T_2,\, \widetilde{T}_1]\) \;=\;& \(2\,i\,J_3\)
		& \(\,[K_3,\, J_3]\) \;=\;&\(0\) &&
		\\[6pt]
		\(\,[T_2,\, \widetilde{T}_2]\) \;=\;& \(-\,2\,i\,K_3\)
		&&&&
		\\
		\hline
	\end{tabular}
	\caption{Commutation relations in the transformed Lorentz algebra (isomorphic to the original).}
	\label{autoLoralgebratable}
\end{table}

For convenience, from this point onward, we relabel the generators
\[
X_1 \;=\; T_1, \quad
X_2 \;=\; T_2, \quad
X_3 \;=\; J_3, \quad
X_4 \;=\; K_3, \quad
X_5 \;=\; \widetilde{T}_1, \quad
X_6 \;=\; \widetilde{T}_2,
\]
so that \(\{\,X_1,\ldots,X_6\}\) is simply another basis for the Lorentz algebra. With the Lorentz algebra recast in this form, we next perform a Lie algebra contraction. Recalling the normal-form approach (cf. Eqs.~\eqref{Y1}--\eqref{Y2} in the previous section), we split the six generators into two sectors:
\[
\underbrace{(X_{1},\,X_{2},\,X_{3},\,X_{4})}_{\text{``first sector''}}
\quad \text{and} \quad
\underbrace{(X_{5},\,X_{6})}_{\text{``second sector''}}.
\]
Hence, the contraction {\it Ansatz} reads
\begin{align}
Y_{1\mu} \;=\; X_{1\mu}
\;+\;
\epsilon \,\sum_{\nu=1}^4 v_{\mu}^{\ \nu}\,X_{1\nu},
\qquad
\mu,\nu \;=\;1,\,2,\,3,\,4,
\label{eq:SIM2Y1}
\\
Y_{2\lambda} \;=\; \epsilon \, X_{2\lambda},
\phantom{aaaaaaaaaaaaaaaaaaaaa....\!} 
\lambda \;=\; 5,\,6.
\label{eq:SIM2Y2}
\end{align}
where the object \(v_\mu^{\ \nu}\) is a completely arbitrary \(4 \times 4\) matrix. Physically, one interprets the first set of generators, \(\{Y_{1\mu}\}\), as those forming the desired \(\mathrm{SIM}(2)\) subgroup in the contraction limit, while the second set, \(\{Y_{2\lambda}\}\), forms an abelian (and invariant) subalgebra. From Table~\ref{autoLoralgebratable}, one sees that the commutator of any pair of first-sector generators never yields a second-sector generator. This observation guarantees that the crucial vanishing condition of Eq.~\eqref{convergencecondition} (from the previous section) is satisfied, ensuring the existence of the limit \(\epsilon \to 0\).

Upon applying the contraction prescription in Eqs.~\eqref{eq:SIM2Y1}--\eqref{eq:SIM2Y2}, one directly adapts the findings of Eqs.~\eqref{resultado1}, \eqref{resultado2}, and \eqref{resultado3}. The result is a decomposition of the new algebra into:
\begin{enumerate}
	\item A four-dimensional subalgebra (spanned by \(Y_{1\mu}\), \(\mu=1,\dots,4\)) isomorphic to $SIM(2)$. Concretely, the relevant commutation relations among \(\{T_1,\,T_2,\,J_3,\,K_3\}\) in the limit \(\epsilon \to 0\) are:
	\begin{align*}
	\label{SIM2_comm1}
	[T_1, T_2] \;=\; 0,\phantom{..}
	[T_1, K_3] \;=\; \phantom{-}i\,T_1,\phantom{..}
	[T_2, K_3] \;=\; i\,T_2,
	\\
	[K_3, J_3] \;=\; 0,\phantom{..}
	[T_1, J_3\phantom{.}] \;=\; \!-\,i\,T_2,\phantom{..}
	[T_2, \phantom{.}J_3] \;=\; i\,T_1.
	\end{align*}
	These define the $SIM(2)$ group, which is often encountered in treatments of massless particle representations and parity violation scenarios (among other contexts).
	
	\item A two-dimensional abelian subalgebra (spanned by \(Y_{2\lambda} \equiv \{\widetilde{T}_1,\,\widetilde{T}_2\}\)), decoupled in the \(\epsilon\to 0\) limit. This abelian sector remains invariant under the first four generators due to the structure implied by Eq.~\eqref{resultado2}. Specifically, its limiting commutation relations are
	\begin{align*}
	[T_1, \widetilde{T}_1] \;=\; 0,
	\quad
	[T_1, \widetilde{T}_2] \;=\; 0,\\
	\quad
	[T_2, \widetilde{T}_1] \;=\; 0,
	\quad
	[T_2, \widetilde{T}_2] \;=\; 0,
	\end{align*}
	together with
	\begin{align*}
	[\widetilde{T}_1, K_3] \;=\; i\,\widetilde{T}_1,
	\quad
	[\widetilde{T}_1, J_3] \;=\;\! -\,i\,\widetilde{T}_2,\\
	\quad
	[\widetilde{T}_2, K_3] \;=\;i\,\widetilde{T}_2,
	\quad
	[\widetilde{T}_2, J_3] \;=\;\phantom{ -} i\,\widetilde{T}_1.
	\end{align*}
\end{enumerate}
Therefore, the six-dimensional Lie algebra in the limit \(\epsilon \to 0\) decomposes into (i) a four-dimensional \(\mathrm{SIM}(2)\) subgroup and (ii) a two-dimensional invariant abelian extension. This behavior precisely mirrors the standard contraction picture wherein part of the algebra remains structurally similar to a lower-dimensional subgroup, while the remainder becomes a commutative sector.

To conclude, we analyze and interpret the obtained results. The contraction procedure elucidated here highlights how specific physically relevant subgroups (like $SIM(2)$) naturally emerge from the Lorentz group in limiting regimes. The abelian sector generated by \(\widetilde{T}_1,\,\widetilde{T}_2\) can then be seen as an extra ``translation-like'' or ``internal'' symmetry sector becoming decoupled as one moves away from the original Lorentz structure. It is worth emphasizing that while the conceptual picture of contraction is straightforward --- enforce a singular linear transformation and then keep finite commutators --- it often requires nontrivial algebraic manipulations to achieve a convenient normal form. In many cases, one strategically chooses combinations of boost and rotation generators, as done here, to isolate the subgroup of interest. In the contraction procedure described above, before taking the limit, the generators \(\tilde{T}_1\) and \(\tilde{T}_2\) are used to untangle, so to speak, the basis of the group space, represented by \(T_1, T_2, K_3\) and \(J_3\). This allows the original Lorentz group basis to be recovered despite the automorphism induced by \(\mathcal{A}\). The limiting operation splits the Lorentz algebraic structure into two distinct subgroups, now defined by \(\{\tilde{T}_1, \tilde{T}_2\}\) and \(\{T_1, T_2, K_3, J_3\}\). In summary, while the Galilei algebra emerges as the classical limit of the Lorentz algebra, the $SIM(2)$ algebra arises as a limit in which Lorentz symmetry is broken since the removal of elements \(\tilde{T}_1\) and \(\tilde{T}_2\) make it impossible to `unfold' the group coordinated directions anymore. The structure chosen to represent the Lorentz group has one particular characteristic that we can make a short exact sequence:
\[
\begin{tikzcd}
0 \arrow[r] & T(2) \arrow[r, "f"] & L^{+}_{^{\uparrow}}  \arrow[r, "g"] & SIM(2) \arrow[r] & 0 ,
\end{tikzcd}
\]
with $f$ and $g$ being the natural morphisms and $T(2)$ the group generated by $\{\widetilde{T}_1,\widetilde{T}_2\}$. Finally, looking at the matrix $\mathcal{A}$ previously defined, one can see that the two lower lines may defined in other ways to obtain different bases for $L^{+}_{^{\uparrow}}$. However, independently of the choice of those two last lines, after the contraction procedure, the result will consistently result in $SIM(2)\oplus T(2)$. 

\section{Four-dimensional \texorpdfstring{\(\mathfrak{sim}(2)\)}{} and \texorpdfstring{\(\mathfrak{isim}(2)\)}{} algebra}

This section presents a method for deriving a four-dimensional closed representation of $\mathfrak{isim}(2)$ algebra. We begin by introducing an initial arrangement that led us to a \(4\times4\) antisymmetric matrix built from \(K_3\), \(J_3\), \(T_1\) and \(T_2\) generators. Subsequently, we outline a systematic procedure to derive the \(\mathfrak{isim(2)}\) algebra within this proposed framework. We first cover its homogeneous sector and then extend the argument to the inhomogeneous counterpart. The antisymmetric property comes from the fact that $SIM(2)$ is a Lorentz subgroup, so its representation is expected to preserve some properties acquired from the broader algebraic structure. In the exposition, we first apply the method to the Lorentz group, and later on, we use it to $SIM(2)$.

By introducing two vectors $K_i = (K_1, K_2, K_3)$ and $J_i = (J_1, J_2, J_3)$, the Lorentz algebra may be represented using the three commutation relations
\begin{eqnarray}
\label{KcomK}
[K_i, K_j] &=& -i\epsilon_{ij}^{\phantom{a.}k}J_k,\\
\label{KcomJ}
\left[K_i, J_j\right] &=& i\epsilon_{ij}^{\phantom{a.}k}K_k,\\
\label{JcomJ}
\left[J_i,J_j\right] &=& i\epsilon_{ij}^{\phantom{a.}k}J_k.
\end{eqnarray}
To find a four-dimensional representation one may introduce a $4\times 4$ matrix $M_{\mu\nu}$, with $\mu,\nu = 0, 1, 2, 3$, such that
\begin{eqnarray}
\label{MemtK}
M_{0i} &=& K_i,\\
\label{MemtJ}
M_{ij} &=& \epsilon_{ij}^{\phantom{a.}k}J_k,
\end{eqnarray}
which, in matrix components, is equivalent to
\begin{eqnarray}
\label{defM0i}
M_{0i} &=& (K_1, K_2, K_3),\\
\label{defMij}
M_{ij} &=& 
\begin{pmatrix}
0 & J_3 & -J_2 \\
-J_3 & 0 & J_1 \\
J_2 & -J_1 & 0
\end{pmatrix}.
\end{eqnarray}
Thus, the algebraic relations for the new objects $M_{0i}$ and $M_{ij}$ must be found from the old ones, defined in (\ref{KcomK})-(\ref{JcomJ}). 

For didactic purposes, we compute all Lie brackets, namely $[M_{0b}, M_{0d}]$, $[M_{ab}, M_{cd}]$ and $[M_{0b}, M_{cd}]$. By doing so, certain results related to the four-dimensional representation of $\mathfrak{sim}(2)$ algebra naturally emerge. Starting from (\ref{MemtK}), after using (\ref{KcomK}) and (\ref{MemtJ}), one may write 
\begin{eqnarray}
\label{kcomKmatM}
[M_{0b}, M_{0d}] &=& [K_b, K_d]\nonumber\\
&=& -i\epsilon_{bd}^{\phantom{a.}k}J_k\nonumber\\
&=& -i M_{bd}
\end{eqnarray}
ans similarly 
\begin{eqnarray}
\label{quaseláMcomM}
[M_{ab}, M_{cd}] = i \epsilon_{ab}^{\phantom{a.}i}\epsilon_{cd}^{\phantom{a.}j} M_{ij}. 
\end{eqnarray} The product $\epsilon_{ab}^{\phantom{a.}i}\epsilon_{cd}^{\phantom{a.}j}$ may be expanded as a determinant of a Kronecker delta matrix yielding
\begin{eqnarray}
\label{relaentreL-C}
\epsilon_{ab}^{\phantom{a.}i}\epsilon_{cd}^{\phantom{a.}j}=\delta_{ac} (\delta_{bd} \delta^{ij} - \delta_{b}^{\phantom{a}j} \delta^i_{\phantom{a}d}) - \delta_{ad} (\delta_{bc} \delta^{ij} - \delta_{b}^{\phantom{a}j} \delta^i_{\phantom{a}c}) + \delta_{a}^{\phantom{a}j} (\delta_{bc} \delta^i_{\phantom{a}d} - \delta_{bd} \delta^i_{\phantom{a}c}).
\end{eqnarray}
Therefore, Eq. (\ref{quaseláMcomM}) reads
\begin{eqnarray}
\label{JcomJmatM}
[M_{ab}, M_{cd}] = i\left(\delta_{bc}  \delta^i_{\phantom{a}d}\delta_{a}^{\phantom{a}j}  - \delta_{ac} \delta^i_{\phantom{a}d}\delta_{b}^{\phantom{a}j}- \delta_{bd} \delta^i_{\phantom{a}c}\delta_{a}^{\phantom{a}j} + \delta_{ad}\delta^i_{\phantom{a}c}\delta_{b}^{\phantom{a}j} \right)M_{ij}.
\end{eqnarray}
In the same manner, $[M_{0b},M_{cd}]$ may be transcribed to the form
\begin{eqnarray}
\label{KcomJmatM}
[M_{0b},M_{cd}] = i \left(  \delta_{bc}\delta^j_{\phantom{a}0}\delta^i_{\phantom{a}d} - \delta_{bd}\delta^j_{\phantom{a}0} \delta^i_{\phantom{a}c}\right)M_{ij}.
\end{eqnarray}

As a final step, let us achieve a general closed-form expression for the Lorentz algebra in four dimensions. In other words, we now seek an expression for the structure constant, $(f_{\mu\nu\phantom{\eta\theta}\rho \sigma}^{\phantom{\mu\nu}\eta\theta})_{Lorentz}$, associated to the Lie bracket $[M_{\mu\nu}, M_{\rho\sigma}]$, with
\begin{equation}
M_{\mu\nu} = 
\begin{pmatrix}
0 & K_1 & K_2 & K_3\\
-K_1 & 0 & J_3 & -J_2 \\
-K_2 & -J_3 & 0 & J_1 \\
-K_3 & J_2 & -J_1 & 0
\end{pmatrix}, 
\end{equation}
being the generator basis of the Lorentz algebra, derived from the combination of both definitions (\ref{defM0i}) and (\ref{defMij}). It is worth noting that one may use the relations (\ref{kcomKmatM}), (\ref{KcomJmatM}), and (\ref{JcomJmatM}) as a guideline since it is expected that the generalized structure constant reduces to these three cases under specific index combinations. This approach ensures consistency with the previously established results while extending the framework to a more general form. Replacing Latin with Greek indices accordingly, expression (\ref{JcomJmatM}) becomes
\begin{eqnarray}
[M_{\mu\nu},M_{\rho\sigma}]=i\left(\eta_{\nu \rho}  \delta^\alpha_{\phantom{a}\sigma}\delta_{\mu}^{\phantom{a}\beta}  - \eta_{\mu \rho} \delta^\alpha_{\phantom{a}\sigma}\delta_{\nu}^{\phantom{a}\beta}- \eta_{\nu \sigma} \delta^\alpha_{\phantom{a}\rho}\delta_{\mu}^{\phantom{a}\beta} + \eta_{\mu \sigma}\delta^\alpha_{\phantom{a}\rho}\delta_{\nu}^{\phantom{a}\beta} \right)M_{\alpha\beta}, 
\end{eqnarray}
being immediately clear that this general expression covers all possibilities. To extract the structure constant from above, it is necessary to consider, due to the index contraction with $M_{\alpha\beta}$, the antisymmetric contribution\footnote{Here, $A_{[\mu\nu]}$ denotes the antisymmetrization of $A_{\mu\nu}$, defined explicitly by $A_{[\mu\nu]} = \frac{1}{2}(A_{\mu\nu}-A_{\nu\mu})$.}:
\begin{equation}
[M_{\mu\nu},M_{\rho\sigma}]=i\left(\eta_{\nu \rho}  \delta_{\mu}^{\phantom{a}[\beta}\delta^{\alpha]}_{\phantom{a}\sigma}  - \eta_{\mu \rho} \delta_{\nu}^{\phantom{a}[\beta}\delta^{\alpha]}_{\phantom{a}\sigma}- \eta_{\nu \sigma} \delta_{\mu}^{\phantom{a}[\beta}\delta^{\alpha]}_{\phantom{a}\rho} + \eta_{\mu \sigma}\delta_{\nu}^{\phantom{a}[\beta}\delta^{\alpha]}_{\phantom{a}\rho} \right)M_{\alpha\beta},
\end{equation}
yielding to the well-known Lorentz structure constant,
\begin{equation}
\label{lorentzstrucconstjao}
(f_{\mu\nu\phantom{\alpha\beta}\rho\sigma}^{\phantom{\mu\nu}\alpha\beta})_{Lorentz} \equiv 2(\eta_{\nu \rho}  \delta^{[\alpha}_{\phantom{a}\sigma}\delta_{\mu}^{\phantom{a}\beta]}  - \eta_{\mu \rho} \delta^{[\alpha}_{\phantom{a}\sigma}\delta_{\nu}^{\phantom{a}\beta]}- \eta_{\nu \sigma} \delta^{[\alpha}_{\phantom{a}\rho}\delta_{\mu}^{\phantom{a}\beta]} + \eta_{\mu \sigma}\delta^{[\alpha}_{\phantom{a}\rho}\delta_{\nu}^{\phantom{a}\beta]}), 
\end{equation}
which is an antisymmetric object concerning the pairs ($\mu,\nu$), ($\rho, \sigma$), ($\alpha,\beta$) and ($\mu\nu, \rho\sigma$). 

Although the transposition of the given approach to the $SIM(2)$ requires supplementary procedural steps, the fundamental framework remains unchanged: we first introduce a well-motivated \textit{Ansatz} that establishes an initial configuration for the three-dimensional vector representation. We start defining  
\begin{eqnarray}
\label{Khatsim2}
\hat{K}_i &=& (T_2, -T_1, K_3),\\
\label{Jhatsim2}
\hat{J}_i &=& (T_1, T_2, J_3). 
\end{eqnarray}
Since a closed-form expression for the three-dimensional $\mathfrak{sim}(2)$ Lie brackets is not available from the outset, it is advantageous to reformulate equations (\ref{Khatsim2}) and (\ref{Jhatsim2}) in terms of the conventional Lorentz generator basis as
\begin{eqnarray}
\label{KAYHAT}
\hat{K}_a &=& \zeta_a^{\phantom{a}c}K_c - \alpha_a^{\phantom{a}c}J_c;\\
\label{JAYHAT}
\hat{J}_a &=& \alpha_a^{\phantom{a}c}K_c + \zeta_a^{\phantom{a}c}J_c,
\end{eqnarray}
where the coefficient matrices $\alpha_a^{\phantom{a}c}$ and $\zeta_a^{\phantom{a}c}$ are respectively defined by 
\begin{equation}
\alpha_a^{\phantom{a}c}=
\begin{pmatrix}
1 & 0 & 0 \\
0 & 1 & 0 \\
0 & 0 & 0
\end{pmatrix},\qquad\qquad
\zeta_a^{\phantom{a}c} = 
\begin{pmatrix}
0 & 1 & 0 \\
-1 & 0 & 0 \\
0 & 0 & 1 
\end{pmatrix},
\end{equation}
with Latin indices ranging from $1$ to $3$. In this manner, all Lie brackets ($[\hat{K}_a, \hat{K}_b]$, $[\hat{K}_a,\hat{J}_b]$, and $[\hat{J}_a, \hat{J}_b]$) may be computed in terms of (\ref{KcomK}) -- (\ref{JcomJ}). Starting with $[\hat{K}_a, \hat{K}_b]$, we have
\begin{eqnarray}
[\hat{K}_a, \hat{K}_b]= \zeta_a^{\phantom{a}c}\zeta_b^{\phantom{a}d}[K_c,K_d] - \zeta_a^{\phantom{a}c}\alpha_b^{\phantom{a}d}[K_c,J_d] - \alpha_a^{\phantom{a}c}\zeta_b^{\phantom{a}d}[J_c,K_d] + \alpha_a^{\phantom{a}c}\alpha_b^{\phantom{a}d}[J_c,J_d]\nonumber
\end{eqnarray}
and on account of relations (\ref{KcomK}) -- (\ref{JcomJ}), it may be recast as
\begin{equation}
\label{khatcomkhatquaselá}
[\hat{K}_a, \hat{K}_b] = i(\alpha_a^{\phantom{a}c}\alpha_b^{\phantom{a}d}-\zeta_a^{\phantom{a}c}\zeta_b^{\phantom{a}d})\epsilon_{cd}^{\phantom{aa}k}J_k -i(\zeta_a^{\phantom{a}c}\alpha_b^{\phantom{a}d} + \alpha_a^{\phantom{a}c}\zeta_b^{\phantom{a}d})\epsilon_{cd}^{\phantom{aa}k}K_k.
\end{equation}
By resorting the identity $(\alpha_a^{\phantom{a}c}\alpha_b^{\phantom{a}d}-\zeta_a^{\phantom{a}c}\zeta_b^{\phantom{a}d})\epsilon_{cd}^{\phantom{aa}k} = (\zeta_a^{\phantom{a}c}\alpha_b^{\phantom{a}d} + \alpha_a^{\phantom{a}c}\zeta_b^{\phantom{a}d})\epsilon_{cd}^{\phantom{aa}l}\zeta_l^{\phantom{a}k}$,
the expression (\ref{khatcomkhatquaselá}) reduces to
\begin{eqnarray}
\label{KHATCOMKHAT}
[\hat{K}_a, \hat{K}_b]= -i(\zeta_a^{\phantom{a}c}\alpha_b^{\phantom{a}d} + \alpha_a^{\phantom{a}c}\zeta_b^{\phantom{a}d})\epsilon_{cd}^{\phantom{aa}k}\hat{J}_k.
\end{eqnarray}
The next Lie bracket to be analyzed is  
\begin{eqnarray}
[\hat{K}_a, \hat{J}_b] =i(\zeta_a^{\phantom{a}c}\zeta_{b}^{\phantom{a}d}-\alpha_{a}^{\phantom{a}c}\alpha_b^{\phantom{a}d})\epsilon_{cd}^{\phantom{aa}k}K_k - i (\zeta_{a}^{\phantom{a}c}\alpha_{b}^{\phantom{a}d}+\alpha_a^{\phantom{a}c}\zeta_b^{\phantom{a}d})\epsilon_{cd}^{\phantom{aa}k}J_k,
\end{eqnarray}
which, after implementing both identities
\begin{equation}
\label{Intermezzo:identidadeimportante}
(\zeta_a^{\phantom{a}c}\zeta_b^{\phantom{a}d}-\alpha_a^{\phantom{a}c}\alpha_b^{\phantom{a}d})\epsilon_{cd}^{\phantom{aa}k} = (\zeta_a^{\phantom{a}c}\alpha_b^{\phantom{a}d}+\alpha_a^{\phantom{a}c}\zeta_b^{\phantom{a}d})\epsilon_{cd}^{\phantom{aa}l}\zeta_l^{\phantom{a}k} 
\end{equation}
and
\begin{equation}
(\zeta_a^{\phantom{a}c}\alpha_b^{\phantom{a}d}+\alpha_a^{\phantom{a}c}\zeta_b^{\phantom{a}d})\epsilon_{cd}^{\phantom{aa}k} = (\zeta_a^{\phantom{a}c}\alpha_b^{\phantom{a}d}+\alpha_a^{\phantom{a}c}\zeta_b^{\phantom{a}d})\epsilon_{cd}^{\phantom{aa}l}\alpha_l^{\phantom{a}k},
\end{equation}
gives 
\begin{eqnarray}
\label{KHATCOMJHAT}
[\hat{K}_a, \hat{J}_b]= i(\zeta_a^{\phantom{a}c}\alpha_b^{\phantom{a}d}+\alpha_a^{\phantom{a}c}\zeta_b^{\phantom{a}d})\epsilon_{cd}^{\phantom{aa}k}\hat{K}_k,
\end{eqnarray}
Lastly, by using relation (\ref{Intermezzo:identidadeimportante}), the commutation $[\hat{J}_a, \hat{J}_b]$ yields 
\begin{equation}
\label{JHATCOMJHAT}
[\hat{J}_a, \hat{J}_b] = i(\zeta_a^{\phantom{a}c}\alpha_b^{\phantom{a}d}+\alpha_a^{\phantom{a}c}\zeta_b^{\phantom{a}d})\epsilon_{cd}^{\phantom{aa}k}\hat{J}_k.
\end{equation}
It is significant to observe that expressions (\ref{KHATCOMKHAT}), (\ref{KHATCOMJHAT}), and (\ref{JHATCOMJHAT}) can be interpreted as a slight modification of the standard Lorentz algebra due to the $(\zeta_a^{\phantom{a}c}\alpha_b^{\phantom{a}d}+\alpha_a^{\phantom{a}c}\zeta_b^{\phantom{a}d})$ factor. We shall return to this factor in a moment. By now, equipped with the three-dimensional $\mathfrak{sim}(2)$ algebra, we may introduce the four-dimensional matrix, $\hat{M}_{\mu\nu}$, defined by
\begin{eqnarray}
\label{MemtKHAT}
\hat{M}_{0i} &=& \hat{K}_i,\\
\label{MemTJHAT}
\hat{M}_{ij} &=& \epsilon_{ij}^{\phantom{aa}k}\hat{J}_k,
\end{eqnarray}
that is 
\begin{equation}
\hat{M}_{\mu\nu} = 
\begin{pmatrix}
0 & T_2 & -T_1 & K_3\\
-T_2 & 0 & J_3 & -T_2 \\
T_1 & -J_3 & 0 & T_1 \\
-K_3 & T_2 & -T_1 & 0
\end{pmatrix}.
\end{equation}

To achieve a closed-form to the structure constant associated with the Lie bracket $[\hat{M}_{\mu\nu},\hat{M}_{\rho\sigma}]$, we do need first find the algebra spanned by $\hat{M}_{0i}$ and $\hat{M}_{ij}$. Following an approach analogous to the preceding case, it can be readily verified that
\begin{equation}
[\hat{M}_{0b},\hat{M}_{0d}] = -i(\zeta_b^{\phantom{a}i}\alpha_d^{\phantom{a}j} + \alpha_b^{\phantom{a}i}\zeta_d^{\phantom{a}j})\hat{M}_{ij},
\end{equation}
\begin{equation}
[\hat{M}_{0b},\hat{M}_{cd}] = i\epsilon_{cd}^{\phantom{aa}n}(\zeta_b^{\phantom{a}i}\alpha_n^{\phantom{a}j}+\alpha_b^{\phantom{a}i}\zeta_n^{\phantom{a}j})\epsilon_{ij}^{\phantom{aa}k}\hat{M}_{0k},
\end{equation}
\begin{equation}
\label{MHATCOMMHATINTERESSE}
[\hat{M}_{ab},\hat{M}_{cd}] = i\epsilon_{ab}^{\phantom{aa}e}\epsilon_{cd}^{\phantom{aa}f}(\zeta_e^{\phantom{a}i}\alpha_f^{\phantom{a}j}+\alpha_e^{\phantom{a}i}\zeta_f^{\phantom{a}j})\hat{M}_{ij}.
\end{equation}
The desired four-dimensional generalization is reached after a suitable replacement of $\alpha$ and $\zeta$ matrices. In fact, after the following adaptation 
\begin{equation}
\label{parquecomplementaaprescrição}
\alpha_a^{\phantom{a}c} \mapsto \alpha_\mu^{\phantom{a}\nu} = 
\begin{pmatrix}
0 & 0 & 0 & 0\\
0 & 1 & 0 & 0 \\
0 & 0 & 1 & 0 \\
0 & 0 & 0 & 0
\end{pmatrix},
\quad \quad \zeta_a^{\phantom{a}c} \mapsto \zeta_\mu^{\phantom{a}\nu} = 
\begin{pmatrix}
0 & 0 & 0 & -1\\
0 & 0 & 1 & 0 \\
0 & -1 & 0 & 0 \\
0 & 0 & 0 & -1
\end{pmatrix},   
\end{equation} 
and performing the necessary antisymmetrization, we have 
\begin{eqnarray}
[\hat{M}_{\mu\nu},\hat{M}_{\rho\sigma}] &=& [\eta_{\mu\rho}(\eta_{\nu\sigma}\eta^{[\alpha\beta]}-\delta_\nu^{\phantom{a}[\beta}\delta_\sigma^{\phantom{a}\alpha]}) - \eta_{\mu\sigma}(\eta_{\nu\rho}\eta^{[\alpha\beta]}-\delta_\nu^{\phantom{a}[\beta}\delta_\rho^{\phantom{a}\alpha]})+\nonumber\\ &+&\delta_\mu^{\phantom{a}[\beta}(\eta_{\nu\rho}\delta^{\phantom{a}\alpha]}_{\sigma}-\eta_{\nu\sigma}\delta^{\phantom{a}\alpha]}_{\rho})](\zeta_\alpha^{\phantom{a}[\eta}\alpha_{\beta}^{\phantom{a}\theta]}+\alpha_\alpha^{\phantom{a}[\eta}\zeta_\beta^{\phantom{a}\theta]})\hat{M}_{\eta\theta}, 
\end{eqnarray}
which, by means of $\eta^{[\alpha\beta]}=0$, simplifies to
\begin{eqnarray}
\label{LBFORSIM(2)4D}
[\hat{M}_{\mu\nu},\hat{M}_{\rho\sigma}] &=& [\eta_{\nu\rho}\delta_\mu^{\phantom{a}[\beta}\delta^{\phantom{a}\alpha]}_{\sigma}-\eta_{\mu\rho}\delta_\nu^{\phantom{a}[\beta}\delta_\sigma^{\phantom{a}\alpha]}-\eta_{\nu\sigma}\delta_\mu^{\phantom{a}[\beta}\delta^{\phantom{a}\alpha]}_{\rho} +\nonumber\\ &+&  \eta_{\mu\sigma}\delta_\nu^{\phantom{a}[\beta}\delta_\rho^{\phantom{a}\alpha]}](\zeta_\alpha^{\phantom{a}[\eta}\alpha_{\beta}^{\phantom{a}\theta]}+\alpha_\alpha^{\phantom{a}[\eta}\zeta_\beta^{\phantom{a}\theta]})\hat{M}_{\eta\theta}. 
\end{eqnarray}
This leads to the explicit form for the $SIM(2)$ structure constant given by
\begin{eqnarray}
\label{eq8}
\hat{f}_{\mu\nu\phantom{\eta\theta}\rho\sigma}^{\phantom{\mu\nu}\eta\theta} &=& 2[\eta_{\nu\rho}\delta_\mu^{\phantom{a}[\beta}\delta^{\phantom{a}\alpha]}_{\sigma}-\eta_{\mu\rho}\delta_\nu^{\phantom{a}[\beta}\delta_\sigma^{\phantom{a}\alpha]}-\eta_{\nu\sigma}\delta_\mu^{\phantom{a}[\beta}\delta^{\phantom{a}\alpha]}_{\rho} + \nonumber\\ &+& \eta_{\mu\sigma}\delta_\nu^{\phantom{a}[\beta}\delta_\rho^{\phantom{a}\alpha]}](\zeta_\alpha^{\phantom{a}[\eta}\alpha_{\beta}^{\phantom{a}\theta]}+\alpha_\alpha^{\phantom{a}[\eta}\zeta_\beta^{\phantom{a}\theta]}),
\end{eqnarray}
an object antisymmetric under the exchange of the index pairs ($\mu,\nu$), ($\rho, \sigma$), ($\eta,\theta$) and ($\mu\nu, \rho\sigma$). 

After comparing the result (\ref{eq8}) with equation (\ref{lorentzstrucconstjao}), the former expression may be rewritten as
\begin{equation}\label{ehessa}
\hat{f}_{\mu\nu\phantom{\eta\theta}\rho\sigma}^{\phantom{\mu\nu}\eta\theta} = (f_{\mu\nu\phantom{\alpha\beta}\rho\sigma}^{\phantom{\mu\nu}\alpha\beta})_{Lorentz}(\zeta_\alpha^{\phantom{a}[\eta}\alpha_{\beta}^{\phantom{a}\theta]}+\alpha_\alpha^{\phantom{a}[\eta}\zeta_\beta^{\phantom{a}\theta]}),
\end{equation}
demonstrating that $SIM(2)$ structure constant is derived from the contraction of  $(\zeta_\alpha^{\phantom{a}[\eta}\alpha_{\beta}^{\phantom{a}\theta]}+\alpha_\alpha^{\phantom{a}[\eta}\zeta_\beta^{\phantom{a}\theta]})$ (in $\alpha$ and $\beta$ indexes) on the Lorentz structure constant. In this context, the process effectively operates as if this factor imposes constraints on the Lorentz structure constants, thereby reducing them to the $SIM(2)$ algebraic framework.   

The steps that led to the equation (\ref{LBFORSIM(2)4D}) cannot be replicated in order to obtain the structure constant associated with the inhomogeneous sector of $\mathfrak{isim(2)}$ algebra, marked by the Lie bracket between $\hat{M}_{\mu\nu}$ and the ordinary translation four-vector, $P^{\mu}$. Therefore, a new approach may be presented. Equation (\ref{ehessa}) qualitatively suggests the following result:
    \begin{equation}
        \label{correçãoinhomoporumamulti}
        [\hat{M}_{\mu\nu},P_\sigma] = i(f_{\mu\nu\phantom{\eta}\sigma}^{\phantom{\mu\nu}\eta})_{Lorentz}C_\eta^{\phantom{a}\rho}P_\rho,
    \end{equation}
    which constraints the usual Lorentz structure constant
    \begin{equation}
        (f_{\mu\nu\phantom{\eta}\sigma}^{\phantom{\mu\nu}\eta})_{Lorentz} = \eta_{\mu\sigma}\delta_\nu^{\phantom{a}\eta}-\eta_{\sigma\nu}\delta_\mu^{\phantom{a}\eta},
    \end{equation}
    by introducing an arbitrary matrix $C$. However, within the proposed framework, it is not possible to consistently determine such a matrix in a way that would allow expression (\ref{correçãoinhomoporumamulti}) to recover the $\mathfrak{isim(2)}$ algebra. The solution proposed consists of adding a linear correction from the outset established by (\ref{correçãoinhomoporumamulti}), yielding to
    \begin{equation}
        \label{correçõesfertig}
        [\hat{M}_{\mu\nu},P_\sigma] = i\{(f_{\mu\nu\phantom{\eta}\sigma}^{\phantom{\mu\nu}\eta})_{Lorentz}\hat{\mathrm{C}}_\eta^{\phantom{a}\rho}+\hat{\mathrm{C}}_{\mu\nu\phantom{\rho}\sigma}^{\phantom{\mu\nu}\rho}\}P_\rho.
    \end{equation}
    In this manner, a possible solution for the matrices $\hat{\mathrm{C}}_\eta^{\phantom{a}\rho}$ and $\hat{\mathrm{C}}_{\mu\nu\phantom{\rho}\sigma}^{\phantom{\mu\nu}\rho}$ can be found by manually assigning their entries based on the expected results directly from $\mathfrak{isim}(2)$ algebra. Notably, the described process yields twenty-four constraint equations that could be used to fix both matrix entries without contradictions. In addition, two other constraints may be found after considering the Lie bracket symmetries:
    \begin{enumerate}
        \item taking into account $[\hat{M}_{\alpha\alpha},P_\sigma]=0$, one may obtain $\hat{\mathrm{C}}_{\alpha\alpha\phantom{\eta}\sigma}^{\phantom{\alpha\alpha}\eta}=0$;
        \item by considering $[\hat{M}_{\mu\nu}, P_\sigma]=-[\hat{M}_{\nu\mu},P_{\sigma}]$, one may find $\hat{\mathrm{C}}_{\mu\nu\phantom{\eta}\sigma}^{\phantom{\mu\nu}\eta}= -\hat{\mathrm{C}}_{\nu\mu\phantom{\eta}\sigma}^{\phantom{\nu\mu}\eta}$.
    \end{enumerate}
    It is worth mentioning that there is not a single form to satisfy all the constraint equations, which implies both matrices $\hat{\mathrm{C}}_\eta^{\phantom{a}\rho}$ and $\hat{\mathrm{C}}_{\mu\nu\phantom{\rho}\sigma}^{\phantom{\mu\nu}\rho}$ are not uniquely determined. However, entries of either matrix, when fixed, uniquely determine those of the other.

\section{Analysis for the \texorpdfstring{$SIM(2)$}{} representations}

We start pinpointing key aspects of Bargmann's theory, necessary to explore the $SIM(2)$ case in detail.  

\subsection{Foundational excerpts of Bargmann's theory}

Borrowing the concept of a ray in the description of quantum states, one can construct an operator ray $\mathbf{U} = \{\tau U\}$, with $|\tau| = 1$ and fixed linear operator on a Hilbert space $U$ being a representative of the ray $\mathbf{U}$. From this definition it follows that the product of ray operators representing group elements $a, b, c, \cdots$ of a group $G$, $\mathbf{U_a}\mathbf{U_b} = \mathbf{U_{ab}}$, naturally leads to $U_a U_b = \omega(a,b)U_{ab}$ where $\omega(a,b)$ is a complex and unimodular phase. The function $\omega(a,b)$ is called local factor. The phase freedom in the mathematical description also allows the writing of a new set of operators $U'_a = \phi(a)U_a$, where $|\phi(a)| = 1$ leading to $U'_aU'_b = \omega'(a,b)U'_{ab}$. This freedom suggests the following definition. 
\begin{definition}
	Two local factors $\omega(a,b)$ and $\omega'(a,b)$ defined in neighborhoods $\mathcal{V}$ and $\mathcal{V'}$ are said to be equivalent if
	\begin{equation}
	\label{eq11}
	\omega'(a,b) = \frac{\phi(a)\phi(b)}{\phi(ab)}\omega(a,b)
	\end{equation}
	is valid in $\mathcal{V}_1 \subset (\mathcal{V} \cap \mathcal{V'}). $
\end{definition}
The theory proposed by Bargmann \cite{barg} starts by analyzing local factors and takes shape when applied to Lie groups relevant to physics. Simply stated, in cases where the transformations are such that $\omega = 1$, the representation is genuine and projective otherwise.   
\begin{definition}
	A ray representation of a group $G$ is continuous if for all $a \in G$, $\mathbf{\Psi}$ belonging to a Hilbert space and $\epsilon \in \mathbb{R^*_ {+}}$ there is a neighborhood $\mathcal{V}$ of $G$ in which the distance between $\mathbf{U_a\Psi}$ and $\mathbf{U_b\Psi}$ is bounded by $\epsilon$ if $b \in \mathcal{V}$.
\end{definition}
Regarding continuity, there are profound implications in the construction arising from Wigner's theorem \cite{wig1}, which allows the selection of the so-called admissible representatives endowed with strong continuity (in the sense of the previous definition). This characteristic culminates in the continuity of local factors\footnote{The proof of this theorem can be seen in \cite{barg} and in addition, there are also more details about these aspects in \cite{JG}.}. 

Let $\{U_a\}$ be a set of admissible representatives of a continuous group $G$ in a given neighborhood $\mathcal{V}$ of $e \in G$, such that $U_e = \mathbb{1}$, and $a, b $ and $ab$ belong to $\mathcal{V}$. Thus, $U_{ab}$ is well defined and belongs to the same ray as $U_aU_b$, giving $U_aU_b = \omega(a,b)U_{ab }$. Clearly, $\omega(e,e) = 1$. Furthermore, the associativity law $(U_aU_b)U_c =   U_a(U_bU_c)$, gives
\begin{equation}
\label{eq12}
\omega(a,b)\omega(ab,c) = \omega(b,c)\omega(a,bc).
\end{equation}
This last expression allows one to define local factors formally.
\begin{definition}
	Every continuous and unimodular complex function $\omega(a,b)$ defined for $a,b \in \mathcal{V}$ is a local factor of $G$ defined on $\mathcal{V}$, if $\omega (e,e) = 1$ and (\ref{eq12}) is valid whenever $ab$ and $bc$ belongs to $\mathcal{V}$.
\end{definition}
Furthermore, if $\mathcal{V}$ coincides with the entire group $\omega$, it is said to be a factor of $G$. Locally, it is also possible to write $\omega(a,b) = e^{i\xi(a,b)}$, where $\xi(a,b) \in \mathbb{R}$, such that the previous definition may be rephrased.
\begin{definition}
	A local exponent of a group $G$ defined in a neighborhood of the origin $\mathcal{V}$ is a real and continuous function $\xi(a,b)$ defined for all $a,b \in \mathcal{V }$ and satisfying:
	\begin{itemize}
		\item $\xi(e,e) = 0$, so that $\omega(e,e)=1$;
		\item $\xi(a,b) + \xi(ab,c) = \xi(b,c) + \xi(a,bc)$, so that (\ref{eq12}) is satisfied.
	\end{itemize}
	If $\mathcal{V}$ coincides with $G$, $\xi$ is said to be an exponent of $G$.
\end{definition}
In addition to this last definition, when making $\omega = e^{i\xi}$ and $\phi(a) = e^{ix(a)}$ in equation (\ref{eq11}), we have
\begin{equation}
\label{eq13}
\xi'(a,b) = \xi(a,b) + x(a) + x(b) - x(ab) = \xi(a,b) + \Delta_{a,b}[x]
\end{equation}
entailing the following helpful definition:
\begin{definition}
	Two local exponents $\xi$ and $\xi'$ defined in $\mathcal{V}$ and $\mathcal{V'}$, respectively, are said to be equivalent if (\ref{eq13}) is valid in a neighborhood $ \mathcal{V}_1 \subset (\mathcal{V} \cap \mathcal{V}')$, where $x(a)$ is a real function defined in $\mathcal{V}^2$ (a neighborhood encompassing products $ab$).
\end{definition}

The concepts presented and defined in Bargmann's work can also be generalized in several aspects: i) the phases can be endowed with differentiability (for Lie groups) through the so-called Iwasawa construction \cite{IWASAWA} (see also \cite{eej} for a step by step account), ii) for connected and simply connected groups, results about continuity can be extended through the whole group \cite{PONTRJAGIN}. Going further, operators that belong to a given set of admissible representatives can be denoted by $e^{i\theta}U_a$ with $\theta \in \mathbb{R}$, so that the representation relation can be seen as 
\begin{equation}
\label{eq14}
(e^{i\theta}U_a)(e^{i\theta'}U_b) = e^{i(\theta + \theta')}\omega(a,b)U_{ab} = e^{i (\theta+\theta'+\xi(a,b))}U_{ab},
\end{equation}
suggesting the following definition:
\begin{definition}
	Let $\xi$ be a local exponent of $G$ defined in $\mathcal{V}$. The local group $H$ is formed by the pairs $\{\theta, a\}$, with $\theta \in \mathbb{R}$ and $a \in \mathcal{V}^2$, with the following composition law
	\begin{equation}
	\label{eq15}
	\{\theta, a\}\cdot\{\theta', b\} = \{\theta + \theta' + \xi(a,b), ab\}.
	\end{equation}
\end{definition}
For this group $H$ the identity and inverse elements are $e_H = \{0,e\}$ and $\{\theta, a\}^{-1} = \{-(\theta + \xi(a, a^{ -1}), a^{-1}\}$, respectively. A relevant property concerning this group definition is the existence of a one-parameter subgroup formed by elements $\{\theta, e\}$ belonging to the center of $H$. Removing it, when possible, is, therefore, removing phases and making the representation genuine.

Consider the Lie algebra $\mathfrak{h}$ of $H$ formed by vectors $\bar{a} = \{\alpha^0, a\}$, with $a = (\alpha^1, \cdots, \alpha^ n)$. The commutator between two elements may be expressed by (see \cite{barg} and \cite{mlc} for details)   
\begin{equation}
\label{eq18}
[\bar{a}, \bar{b}] = \{\Xi(a,b),[a,b]\},
\end{equation} where 
\begin{equation}
\label{eq17}
\Xi(a,b) = \lim_{\mu\to 0} \mu^{-2}\Bigg[\xi\big((\mu a,\mu b),(\mu a)^{-1}(\mu b)^{-1}\big) + \xi(\mu a,\mu b) + \xi(-\mu a,-\mu b)\Bigg],
\end{equation} a bilinear antisymmetric form. For the local group, the Jacobi identity takes the form $\bar{j}=\{d\Xi(a,a',a''),j)\}$, where $ d\Xi(a,a' ,a'') = \Xi([a,a'],a'')+\Xi([a',a''],a)+\Xi([a'',a],a') $ and $j = \big[[a,a'],a''\big]+\big[[a',a''],
a\big] + \big[[a'',a],a'\big] = 0$. The following definition entails the prominence of $\Xi$ for Bargmann's approach.
\begin{definition}
	Any real-valued antisymmetric bilinear form $\Xi$, defined in the Lie algebra $\mathfrak{g}$, for which $d\Xi(a,a',a'')=0$ will be called infinitesimal exponent of $\mathfrak{g}$. If the infinitesimal exponent is given by (\ref{eq17}) it is said to correspond to the canonical local exponent $\xi$ of $G$.
\end{definition} Since $[\bar{a}, \bar{b}]:H\times H\rightarrow H$ we can rewrite it in the form $\{\Xi(a,b), e\}\cdot\{ 0, [a,b ]\}$, where $\{\Xi(a,b),e\}$ belongs to the algebra center. This fact enables one to recognize infinitesimal exponents as central charges \cite{WEINBERG}, another well-known method for exploring projective representations. Infinitesimal exponents may also have associated an equivalence relation. This is the point formalized by the next definition.   
\begin{definition}
	Two infinitesimal exponents $\Xi$ and $\Xi'$ of $\mathfrak{g}$ are said to be equivalent if $\Xi'(a,b) = \Xi(a,b) -\Lambda([a,b])$ and we can write $\Xi'\equiv \Xi$, where $\Lambda(a)$ is a real linear form defined in $\mathfrak{g}$.
\end{definition}

From this point on, Bargmann's method consists of manipulating the algebraic relations using Jacobi identities to search for equivalence relations that may be used to conclude, provided this is the case, that a given exponent is equivalent to zero. Such a process can successfully demonstrate that Lorentz and Poincaré groups in more than two dimensions have a genuine local representation, and the Galilei group has a locally projective representation.  

In the sequel, we shall apply the method to the $SIM(2)$ group and evince a caveat of the formalism: for this group, there is a commutation relation that is not reached through an algebraic rearrangement via Jacobi identities. Consequently, no conclusion can be inferred about the corresponding infinitesimal exponent. More specifically, all the phases can be removed except, possibly, the phase associated with the aforementioned infinitesimal exponent.   
\subsection{\texorpdfstring{$SIM(2)$}{} (and \texorpdfstring{$ISIM(2)$}{}) group representation}

Starting from the Lie bracket defined by (\ref{LBFORSIM(2)4D}), it is readily obtained 
\begin{eqnarray}
[\hat{M}_{\lambda}^{\phantom{a}\kappa}, \hat{M}_{\kappa\gamma}]=i \Sigma^{\eta \theta}_{\phantom{aa}\gamma\lambda} \hat{M}_{\eta\theta},
\end{eqnarray}
where 
\begin{equation}
\label{eq4}
\Sigma^{\eta \theta}_{\phantom{aa}\gamma\lambda} = 2(\zeta_{[\gamma}^{\phantom{a}[\eta}\alpha_{\lambda]}^{\phantom{a}\theta]}+ \alpha_{[\gamma}^{\phantom{a}[\eta}\zeta_{\lambda]}^{\phantom{a}\theta]}).
\end{equation}
Therefore, the computation of the bilinear antisymmetric form $\Xi([\hat{M}_{\lambda}^{\phantom{a}\kappa}, \hat{M}_{\kappa\gamma}], \hat{M}_{\rho\sigma} )$, gives
\begin{equation}
\label{eq1}
\Xi([\hat{M}_{\lambda}^{\phantom{a}\kappa}, \hat{M}_{\kappa\gamma}], \hat{M}_{\rho\sigma} )  = i\Sigma^{\eta \theta}_{\phantom{aa}\gamma\lambda}\Xi(\hat{M}_{\eta\theta}, \hat{M}_{\rho\sigma}). 
\end{equation}
The inside $\Xi$ Jacobi identity $\Xi([\hat{M}_{\lambda}^{\phantom{a}\kappa}, \hat{M}_{\kappa\gamma}], \hat{M}_{\rho\sigma} ) + \Xi([\hat{M}_{\kappa\gamma}, \hat{M}_{\rho\sigma}], \hat{M}_{\lambda}^{\phantom{a}\kappa}) + \Xi([\hat{M}_{\rho\sigma}, \hat{M}_{\lambda}^{\phantom{a}\kappa}], \hat{M}_{\kappa\gamma}) = 0$, can be used to rewrite Eq. (\ref{eq1}) as 
\begin{eqnarray}
\label{eq2}
\Xi(\hat{M}_{\lambda}^{\phantom{a}\kappa},[\hat{M}_{\kappa\gamma}, \hat{M}_{\rho\sigma}]) + \Xi(\hat{M}_{\kappa\gamma}, [\hat{M}_{\rho\sigma}, \hat{M}_{\lambda}^{\phantom{a}\kappa}]) &=& i\Sigma^{\eta\theta}_{\phantom{aa}\gamma\lambda}\Xi(\hat{M}_{\eta\theta}, \hat{M}_{\rho\sigma}),
\end{eqnarray}
Now, from the Lie brackets relation, the left-hand side of (\ref{eq2}) becomes
\begin{equation}
\label{eq3}
\Sigma^{\mu\nu}_{\phantom{\mu\nu}\gamma\lambda}\Xi(\hat{M}_{\mu\nu}, \hat{M}_{\rho\sigma}) = \hat{f}^{\kappa\phantom{\gamma}\eta\theta}_{\phantom{\kappa}\gamma\phantom{\eta\theta}\rho\sigma}\Xi(\hat{M}_{\lambda\kappa},\hat{M}_{\eta\theta})-\hat{f}^{\kappa\phantom{\lambda}\eta\theta}_{\phantom{\kappa}\lambda\phantom{\eta\theta}\rho\sigma}\Xi(\hat{M}_{\gamma\kappa},\hat{M}_{\eta\theta}).
\end{equation}

Since the left-hand side of Eq. (\ref{eq3}) is partially contracted, it is not straightforward to use the standard Bargmann protocol by isolating $\Xi(\hat{M}_{\mu\nu}, \hat{M}_{\rho\sigma})$ and defining a linear application of the type $\Lambda(\Sigma^{-1}\hat{f}\Xi)$ such that $\Lambda=\Xi$, rendering null phases for all cases \cite{barg}. Hence, it is not clear at this point whether the standard procedure can be further pursued. Nevertheless, it is possible to push formalism in a specific sense by attempting to find particular infinitesimal exponents. In this regard, let us explore the possibility of finding an exponent containing $J_3 = \hat{M}_{12}$ and $K_3 = \hat{M}_{03}$ (or $J_3 = -\hat{M}_{21}$ and $K_3 = -\hat{M}_{03}$). The left-hand side of (\ref{eq3}) gives
\begin{eqnarray}
\Sigma^{\mu\nu}_{\phantom{aa}\gamma\lambda}\Xi(\hat{M}_{\mu\nu}, \hat{M}_{\rho\sigma}) &=& \Sigma^{00}_{\phantom{aa}\gamma\lambda}\Xi(\hat{M}_{00}, \hat{M}_{\rho\sigma}) + \Sigma^{i0}_{\phantom{aa}\gamma\lambda}\Xi(\hat{M}_{i0}, \hat{M}_{\rho\sigma}) +\nonumber\\ &+& \Sigma^{0j}_{\phantom{aa}\gamma\lambda}\Xi(\hat{M}_{0j}, \hat{M}_{\rho\sigma}) + \Sigma^{ij}_{\phantom{aa}\gamma\lambda}\Xi(\hat{M}_{ij}, \hat{M}_{\rho\sigma}),
\end{eqnarray}
which after considering $\Sigma^{0i}_{\phantom{aa}\gamma\lambda} =  -\Sigma^{i0}_{\phantom{aa}\gamma\lambda}$ together with antisymmetric character of $\hat{M}_{\mu\nu}$, reduces to
\begin{equation}
\Sigma^{\mu\nu}_{\phantom{aa}\gamma\lambda}\Xi(\hat{M}_{\mu\nu}, \hat{M}_{\rho\sigma}) = 2\Sigma^{i0}_{\phantom{aa}\gamma\lambda}\Xi(\hat{M}_{i0}, \hat{M}_{\rho\sigma}) + \Sigma^{ij}_{\phantom{aa}\gamma\lambda}\Xi(\hat{M}_{ij}, \hat{M}_{\rho\sigma}).
\end{equation}
Using (\ref{eq4}) and $\alpha_\mu^{\phantom{a}0} = 0 = \zeta_\mu^{\phantom{a}0}$ for any value of the $\mu$ index, one can show that $\Sigma^{i0}_{\phantom{aa}\gamma\lambda}= \zeta_{[\gamma}^{\phantom{a}i}\alpha_{\lambda]}^{\phantom{a}0}+\alpha_{[\gamma}^{\phantom{a}i}\zeta_{\lambda]}^{\phantom{a}0} - \zeta_{[\gamma}^{\phantom{a}0}\alpha_{\lambda]}^{\phantom{a}i}-\alpha_{[\gamma}^{\phantom{a}0}\zeta_{\lambda]}^{\phantom{a}i}
$ vanishes, leading to
\begin{equation}\label{nnn}
\Sigma^{\mu\nu}_{\phantom{aa}\gamma\lambda}\Xi(\hat{M}_{\mu\nu}, \hat{M}_{\rho\sigma}) = \Sigma^{ij}_{\phantom{aa}\gamma\lambda}\Xi(\hat{M}_{ij}, \hat{M}_{\rho\sigma}).
\end{equation}
The only possibility for the appearance of any term of interest arises from the particular combination $(\rho\sigma) = (03)$. Delving deeper into the analysis, we can examine all possible configurations resulting from the summation over the index pair $(ij)$:
\begin{eqnarray}
\label{eq5}
\Sigma^{ij}_{\phantom{aa}\gamma\lambda}\Xi(\hat{M}_{ij}, \hat{M}_{03}) = 2\Sigma^{12}_{\phantom{aa}\gamma\lambda}\Xi(\hat{M}_{12}, \hat{M}_{03}) + 2\Sigma^{13}_{\phantom{aa}\gamma\lambda}\Xi(\hat{M}_{13}, \hat{M}_{03}) + 2\Sigma^{23}_{\phantom{aa}\gamma\lambda}\Xi(\hat{M}_{23}, \hat{M}_{03}).
\end{eqnarray}
Note that on account of $\Sigma^{ij}_{\phantom{aa}\gamma\lambda} = -\Sigma^{ji}_{\phantom{aa}\gamma\lambda}$, entrances living on the diagonal, like $\Sigma^{ii}_{\phantom{aa}\gamma\lambda}$, are identically zero. Among all terms on the right-hand side of (\ref{eq5}), the significant one is $2\Sigma^{12}_{\phantom{aa}\gamma\lambda}\Xi(\hat{M}_{12}, \hat{M}_{03})$, which should be examined in more detail. By resorting relation (\ref{eq4}), we arrive at
\begin{eqnarray}
\label{eq6}
\Sigma^{12}_{\phantom{aa}\gamma\lambda} &=& \zeta_{[\gamma}^{\phantom{a}1}\alpha_{\lambda]}^{\phantom{a}2}+\alpha_{[\gamma}^{\phantom{a}1}\zeta_{\lambda]}^{\phantom{a}2} - \zeta_{[\gamma}^{\phantom{a}2}\alpha_{\lambda]}^{\phantom{a}1}-\alpha_{[\gamma}^{\phantom{a}2}\zeta_{\lambda]}^{\phantom{a}1}
\end{eqnarray}
and verifying all ($\gamma\lambda$) combinations is necessary. It can be readily seen that if $\gamma = \lambda$ or if any of both indices turns out to be equal to $0$ or $3$, the right-hand side of (\ref{eq6}) is automatically null. The remaining combinations are $(\gamma\lambda) = (12)$ and $(\gamma\lambda) = (21)$, which also vanishes identically. Thus, there is no occurrence of $\Xi(J_3,K_3)$ on the left side of (\ref{eq3}). In what follows, we shall prove that this is also the case for the right-hand side of (\ref{eq3}). 

From (\ref{eq3}), it is evident that contributions proportional to $\Xi(J_3,K_3)$ can only occur if the index pair ($\eta \theta$) takes on combinations such as ($03$) or ($12$), which, on its turn, is impossible due to the inherent algebraic structure of $SIM(2)$. If this scenario were feasible, it would imply the existence of a non-zero structure constant for which the Lie bracket between any group generators could yield $K_3$ or $J_3$, and this is not the case. Aiming completeness of the demonstration, we then prove that $\hat{f}_{\mu\nu\phantom{03}\rho\sigma}^{\phantom{\mu\nu}03} = 0 = \hat{f}_{\mu\nu\phantom{12}\rho\sigma}^{\phantom{\mu\nu}12} $ holds for any $\mu,\nu,\rho$ and $\sigma$. Starting by $(\eta\theta) = (03)$, using the definitions of $\alpha_\mu^{\phantom{a}\alpha}$ and $\zeta_\mu^{\phantom{a}\alpha}$, one may easily check that the right-hand side of (\ref{eq8}) vanishes immediately. Although the $(\eta\theta) = (12)$ case verification process is analogous, it is nevertheless indirect. Expression (\ref{eq8}) become
\begin{eqnarray}
\label{eq10}
\hat{f}_{\mu\nu\phantom{12}\rho\sigma}^{\phantom{\mu\nu}12} &=& 2[\eta_{\nu\rho}(\zeta_{[\sigma}^{\phantom{a}[1}\alpha_{\mu]}^{\phantom{a}2]}+\alpha_{[\sigma}^{\phantom{a}[1}\zeta_{\mu]}^{\phantom{a}2]})-\eta_{\mu\rho}(\zeta_{[\sigma}^{\phantom{a}[1}\alpha_{\nu]}^{\phantom{a}2]}+\alpha_{[\sigma}^{\phantom{a}[1}\zeta_{\nu]}^{\phantom{a}2]})-\nonumber\\ &-& \eta_{\nu\sigma}(\zeta_{[\rho}^{\phantom{a}[1}\alpha_{\mu]}^{\phantom{a}2]}+\alpha_{[\rho}^{\phantom{a}[1}\zeta_{\mu]}^{\phantom{a}2]}) + \eta_{\mu\sigma}(\zeta_{[\rho}^{\phantom{a}[1}\alpha_{\nu]}^{\phantom{a}2]}+\alpha_{[\rho}^{\phantom{a}[1}\zeta_{\nu]}^{\phantom{a}2]})]
\end{eqnarray}
and all the work now comes down to demonstrating that the combination
\begin{equation}
\label{eq9}
(\zeta_{[\sigma}^{\phantom{a}[1}\alpha_{\mu]}^{\phantom{a}2]}+\alpha_{[\sigma}^{\phantom{a}[1}\zeta_{\mu]}^{\phantom{a}2]}) = \zeta_{[\sigma}^{\phantom{a}1}\alpha_{\mu]}^{\phantom{a}2}+\alpha_{[\sigma}^{\phantom{a}1}\zeta_{\mu]}^{\phantom{a}2} - \zeta_{[\sigma}^{\phantom{a}2}\alpha_{\mu]}^{\phantom{a}1} - \alpha_{[\sigma}^{\phantom{a}2}\zeta_{\mu]}^{\phantom{a}1},
\end{equation}
reduces to zero independently of $\sigma$ and $\mu$ values. Again, for all possible combinations of ($\sigma \mu$), all right-hand side terms of (\ref{eq9}) readily vanish. Thus, from Eq. (\ref{eq10}) follows the result $\hat{f}_{\mu\nu\phantom{12}\rho\sigma}^{\phantom{\mu\nu}12} = 0$ for all indexes. In conclusion, the infinitesimal exponent $\Xi(K_3, J_3)$ does never appear in either sides of (\ref{eq3}). 

It must be stressed that the absence of an exponent relating $K_3$ and $J_3$ could be thought of as a result of $\alpha$, $\zeta$, and $\hat{M}$ particular choices. However, the inaccessibility of $\Xi(K_3,J_3)$ can be obtained by direct computation. To exemplify the standard procedure, consider 	$\Xi([T_1, K_3], T_2) = i\Xi(T_1, T_2)$. The Jacobi identity allows one to write $-\Xi([T_2, T_1], K_3)-\Xi([K_3, T_2], T_1) = i\Xi(T_1, T_2)$ and taking into account that $[T_1,T_2]=0$ and $[T_2,K_3]=iT_2$, we are left with $\Xi(T_1,T_2)=0$, provided $\Xi$ is antisymmetric. No phase is expected to survive from this infinitesimal exponent. Another typical case is exemplified, for instance, by $\Xi([T_2, K_3], J_3) = i\Xi(T_2, J_3)$. Proceeding as before we arrive at $\Xi(T_1, J_3)=-\Xi(T_2, K_3)$. This last constraint can be further explored: calling $\Xi([T_1,K_3],J_3)=i\kappa$ and defining the linear form action $\Lambda$ in the Lie algebra such that $\Lambda([T_1,K_3])=\Lambda(iT_1)=i\Lambda(T_1):=i\kappa$, it is trivial to see that $\Xi(T_1,J_3)=\Lambda(T_1)\equiv 0$, a result also inherited by $\Xi(T_2,K_3)$. Hence, no phase is associated with these infinitesimal exponents, since they are equivalent to zero. The point to be stressed here is that, since any commutation relation cannot achieve $K_3$ and $J_3$, the corresponding exponent $\Xi(K_3, J_3)$ is not accessed by the formalism and all that can be asserted is that the phases associated to projective representation are all locally equivalent to one, except, possibly, the one associated to $\Xi(K_3, J_3)$.   

We shall comment on the local and global aspects of this analysis. By now, we stress that, as mentioned in the main text, the study performed here is {\it pari passu} with the standard procedure of removing central charges. Rephrasing this section's findings in the central charges language, we start by adding them to the algebraic relations as dictates table (\ref{tab:asim}).
\begin{table}[ht]
	\centering
	\begin{tabular}{llr|llr} 
		\hline
		\\
		$[T_1, T_2]$ &$=$& $iC(T_1, T_2)$ & $[T_1, K_3]$ &$=$& $iT_1 + iC(T_1, K_3)$\\
		
		$[T_1, J_3]$ &$=$& $-iT_2 + iC(T_1, J_3)$ &$[T_2, K_3]$ &$=$& $iT_2 + iC(T_2, K_3)$\\ 
		
		$[T_2, J_3]$ &$=$& $iT_1 + iC(T_2, J_3)$ &$[J_3, K_3]$ &$=$& $iC(J_3, K_3)$\\
		\\
		\hline
	\end{tabular}
	\caption{Standard $\mathfrak{sim(2)}$ algebra with central charges.}
	\label{tab:asim}
\end{table}
A bit of simple but tedious algebra evinces the following conclusions: the Jacobi identity for the $(T_1,T_2, K_3)$ generators leads to $C(T_1, T_2) = 0$, while for $(T_1, T_2, J_3)$ does not provide information. For generators $(T_1,K_3,J_3)$, this procedure leads to $C(T_1, J_3) = -C(T_2, K_3)$ and for the set $(T_2,K_3,J_3)$ of generators we are left with $C(J_3, T_2) = C(K_3, T_1)$. Observe that these are nothing but relations obtained through manipulating infinitesimal exponents. The generator redefinition $\bar{T}_1 = T_1 + C(T_1, K_3)$ and $\bar{T}_2 = T_2 + C(T_2, K_3)$ automatically gives standard algebraic relations (see table (\ref{tab:asimr})), but with $C(J_3,K_3)$ as a residual central charge.     
\begin{table}[ht]
	\centering
	\begin{tabular}{llr|llr}
		\hline
		\\
		$[\bar{T}_1, \bar{T}_2]$ &$=$& $0$ & $[\bar{T}_1, K_3]$ &$=$& $i\bar{T}_1$\\
		
		$[\bar{T}_1, J_3]$ &$=$& $-i\bar{T}_2$ &$[\bar{T}_2, K_3]$ &$=$& $i\bar{T}_2$\\ 
		
		$[\bar{T}_2, J_3]$ &$=$& $i\bar{T}_1$ &$[J_3, K_3]$ &$=$& $iC(J_3, K_3)$\\
		\\
		\hline
	\end{tabular}
	\caption{$\mathfrak{sim(2)}$ algebra. Note the $C(J_3,K_3)$ central charge presence.}
	\label{tab:asimr}
\end{table}

Incidentally, it is emphasized that the associativity law corresponding version of a local exponent (presented in Definition IV) may be faced as a specific coboundary. If this coboundary is not a cocycle, it defines an element $[h]$ of the second group cohomology \cite{mlc}. Such an element would have as representative a map $h:SIM(2)\times SIM(2)\rightarrow S^1$. Therefore, the existence of projective representations (locally) is associated with $H^2_{gr}(SIM(2),S^1)$ \cite{brown,corry}. The relevant group sector is $SIM(2)\supset SO(2)\cong SPIN(2)$, which has a topological structure of $S^1$. Thus, the extensions measured by $H^2_{gr}$ are given essentially by a relation whose net effect is the map $S^1\rightarrow S^1$. Therefore, $H^2_{gr}(SIM(2),S^1)\cong \mathbb{Z}$. Translations can be encompassed to $SIM(2)$ transformations via a usual semisimple extension, in which the resulting group is denoted by $ISIM(2)$. Due to the coboundary operator differential nature, it is expected that for $H^2_{gr}(ISIM(1),S^1)$, the same nontriviality of its homogeneous counterpart. This is the case, in fact: resorting to the analysis of central extensions for $\mathfrak{isim(2)}$, the generators redefinition performed for the homogeneous case along with the following momentum redefinition $P_0\mapsto P_0 + C(K_3, P_3)$, $P_1\mapsto P_1 + C(J_3, P_2)$, $P_2\mapsto P_2 + C(J_3, P_1)$, and $P_3\mapsto P_3 + C(K_3, P_0)$ restore the usual algebraic relations, but the central charge $C(J_3,K_3)$ is not eliminated (see Table (\ref{lbe})). In terms of cohomology language, $H^2_{gr}(ISIM(1),S^1)\cong\mathbb{Z}$.        

As previously emphasized, all the results so far concern local phases. Globally, by its turn, the analysis is quite similar to that of the Lorentz and Poincaré groups. The topology associated with $SIM(2)$ is $\mathbb{R}^2\times\mathbb{R}^+\times S^1/\mathbb{Z}_2$ and, as such, a sign associated with fermionic representations is also expected. 
\begin{table}[ht]
	\centering
	\begin{tabular}{llr|llr}
		\hline
		\\
		$[T_1, T_2]$ &$=$& $0$ & $[T_1, K_3]$ &$=$& $i T_1$\\
		$[T_1, J_3]$ &$=$& $-i T_2$ &$[T_2, K_3]$ &$=$& $i T_2$\\ 
		$[T_2, J_3]$ &$=$& $i T_1$ &$[J_3, K_3]$ &$=$& $iC(J_3, K_3)$\\
		$[P_\mu, P_\nu]$ &$=$& $0$\\
		\\
		\hline
		\\
		$[T_1, P_0]$ &$=$& $ i P_1$ & $[T_2, P_0]$ &$=$& $i P_2$\\
		$[T_1, P_1]$ &$=$& $i P_0 - i P_3$ &$[T_2, P_1]$ &$=$& $0$\\ 
		$[T_1, P_2]$ &$=$& $0$ &$[T_2, P_2]$ &$=$& $i P_0 - i P_3$\\
		$[T_1, P_3]$ &$=$& $i P_1$ &$[T_2, P_3]$ &$=$& $i P_2$\\
		\\
		\hline
		\\
		$[K_3, P_0]$ &$=$& $ i P_3$ & $[J_3, P_0]$ &$=$& $0$\\
		$[K_3, P_1]$ &$=$& $0$ &$[J_3, P_1]$ &$=$& $i P_2$\\ 
		$[K_3, P_2]$ &$=$& $0$ &$[J_3, P_2]$ &$=$& $-i P_1$\\
		$[K_3, P_3]$ &$=$& $i P_0$ &$[J_3, P_3]$ &$=$& $0$\\
		\\
		\hline
	\end{tabular}
	\caption{$\mathfrak{isim(2)}$ algebra (omitting the bar). Note the $C(J_3,K_3)$ central charge presence.} 
	\label{lbe}
\end{table}
We have said that Bargmann's formalism is still particularly useful for tracking the specific phase source in projective representation. In the next section, we present a straightforward way of determining whether a group representation is genuine or projective, though the method cannot depict the phase source or explicit form in projective representation.
\subsection{A complementary viewpoint for non-abelian groups}
As it is clear from the previous Bargmann theory exposition and application to the $SIM(2)$ group case if a given infinitesimal exponent is not reached through Jacobi identities, it is not possible to relate it to a linear algebraic mapping leading to an equivalence relation of a null exponent. Of course, there are cases for which, even assessing the infinitesimal exponent, there is no equivalence with zero, precisely the cases for which the representation is projective. However, the point is that whenever an infinitesimal exponent is not accessible, the formalism stops being exhaustive. By its turn, a given infinitesimal exponent $\Xi(A,B)$ is not reachable when the generator $A$ (or $B$) is not obtained from any other algebraic commutation relation. The following definition aims to formalize these observations. 

\begin{definition}
	The $R-$set of a generator $b$ of an algebra $\mathfrak{g}$ associated to the non-abelian group $G$ is given by all the elements of $\mathfrak{g}$ accessed by a commutation with $b$:
	\begin{equation}
	\label{401}
	R(b) = \{[b,x]\ | \forall x \in \{\mathfrak{g}\} \}, 
	\end{equation}
    where $\{\mathfrak{g}\}$ denotes the set of $\mathfrak{g}$ generators. 
\end{definition} 

Therefore, whenever an element does not belong to any $R-$set, a projective local phase is expected since its corresponding infinitesimal exponent is unreachable, although these phase factors may originate from distinct commutation relations on each respective case. Let us revisit specific and relevant cases, starting with Pseudo-Ortogonal groups.      

Recall that the tridimensional Poincaré algebra is given by 
\begin{equation}
[J_i,J_j]=i\varepsilon_{ijk}J_k;\ \ \ \ \ \ [J_i,K_j]=i\varepsilon_{ijk}K_k;\ \ \ \ \ \ [K_i,K_j]=-i\varepsilon_{ijk}J_k;\nonumber 
\end{equation}
\begin{equation}
[J_i,P_j]=i\varepsilon_{ijk}P_k;\ \ \ \ \ \ [K_i,P_j]=iH\delta_{ij};\ \ \ \ \ \ [K_i,H]=iP_i;\nonumber    
\end{equation}
\begin{equation}
[P_i,P_j]=[P_i,H]=[J_i,H]=0,\nonumber    
\end{equation} where $\{J_i\}$, $\{K_i\}$, $\{P_i\}$ $(i=1,2,3)$ are the rotations, boosts, and spatial translation generators, respectively, and $H$ stands for the Hamiltonian (the time translating generator). The $R-$sets are given by $R(\vec{J})=\{\vec{J},\vec{K},\vec{P}\}$, $R(\vec{K})=\{\vec{J},\vec{K},\vec{P},H\}$, $R(\vec{P})=\{H,\vec{P}\}$, and $R(H)=\{\vec{P}\}$. Every generator belongs to at least one $ R-$set, and thus, projective representation is not in order. This is the (well-known) case for the Poincaré (and Lorentz) group in more than two dimensions. In $1+1$ dimensions, the situation is slightly different. The three generators are $H$ for temporal translation, $P$ for spacial translation, and $K$ performing a boost along the spacial direction. The commutators are
\begin{equation}
[K,H]=P; \ \ \ \ \ \ [K,P]=H; \ \ \ \ \ \ [H,P]=0.
\end{equation} The $R-$sets read $R(K)=\{H,P\}$, $R(P)=\{H\}$, and $R(H)=\{P\}$. Observe that $K$ does not belong to any $R-$set and, therefore, a phase factor is in order. This is the case for the two-dimensional Poincaré group \cite{barg}.

Going further, let us contrast this point of view with another well-known case presenting projective representation. Consider the Galilei algebra with generators set given by $\{a,b,d,f\}$, where $a$ is associated with orthogonal transformations in 3 dimensions, $b$ is associated with pure translations, $d$ is responsible for Galilean boosts, and $f$ performs temporal translation. This set of generators satisfies  
\begin{equation}
\label{421}
[a_{ij},a_{kl}] = \delta_{jk}a_{il} - \delta_{ik}a_{jl} + \delta_{il}a_{jk} - \delta_{jl}a_{ik};
\end{equation}
\begin{equation}
\label{422}
[a_{ij}, b_k] =  \delta_{jk}b_i - \delta_{ik}b_j; \ \ \ \ \ \ [b_i, b_j] = 0;
\end{equation}
\begin{equation}
\label{423}
[a_{ij}, d_k] =  \delta_{jk}d_i - \delta_{ik}d_j; \ \ \ \ \ \ [d_i, d_j] = 0; \ \ \ \ \ \ [d_i, b_j] = 0;
\end{equation}
\begin{equation}
\label{424}
[a_{ij}, f] = 0; \ \ \ \ \ \ [b_k, f] = 0; \ \ \ \ \ \ [d_k, f] = b_k.
\end{equation}
The $R-$sets are given by $R(a)=\{a,b,d\}$, $R(b)=\{b\}$, $R(d)=\{b,d\}$, and $R(f)=\{b\}$. It can be readily seen that $f$ does not belong to any $R-$set, and a phase factor is expected, as is the case indeed. 

Returning to the $\mathfrak{sim(2)}$ and its inhomogeneous counterpart, recall that the commutators are given by 
\begin{equation}
[T_1, T_2] = 0; \ \ \ \ \ \ [T_1, K_3] = iT_1; \ \ \ \ \ \ [T_1, J_3] = -iT_2; \nonumber
\end{equation}
\begin{equation}
[T_2, K_3] = iT_2; \ \ \ \ \ \ [T_2, J_3] = iT_1; \ \ \ \ \ \ [K_3, J_3] = 0 \nonumber
\end{equation} and the $R-$sets all coincide to $\{T_1,T_2\}$. As seen, $K_3$ and $J_3$ are not in any $R-$set. Hence, the corresponding infinitesimal exponent is completely inaccessible; again, a phase factor is due. The $R-$sets are changed for the $\mathfrak{isim(2)}$, but $K_3$ and $J_3$ still not belonging to them.  

\section{Final Remarks}

Lorentz group representation, details, and usefulness have found shelter in many branches of high-energy physics. Despite its common usage, the richness of its structure is truly impressive, entering the physical description from relativistic kinematics to fundamental particles. We believe that, at least in part, it is also true for $SIM(2)$. This paper presented several important results concerning the $SIM(2)$ Lorentz subgroup, aiming to furnish a mathematically inclined basis for further physical investigations, presenting relevant results about algebraic and group representations. 

Our explicit construction of the $SIM(2)$ representations may serve as a starting point for obtaining the algebraic tools necessary to investigate integrable models possessing VSR symmetry \cite{bale1,bale2}. Furthermore, our analysis of the cohomological aspects (via Bargmann's formalism) establishes the necessary framework for central extensions, which play a key role in the quantization of two-dimensional integrable field theories \cite{bale2} within the VSR scope.

We finalize our considerations by attempting to frame $SIM(2)$ symmetries in a broad perspective, though we are deliberately entering a speculative realm. Gauge systems are believed to present more symmetry at high energies. That is precisely the case for gauge interactions. Systems are, however, for which the behavior is the opposite, that is: the higher the energy, the lower the symmetry, as occurs in some fluid systems\footnote{Think of the cigarette smoke, for example.}. It would be intriguing if full Lorentz symmetries were acquired only at some energy scale, but for more extreme cases, spacetime would require fewer symmetries, allowing, for instance, for a privileged direction. 
   
\section{Acknowledgements}

JER, JMBM, and GMCR thanks CAPES for for financial support. JMHS thanks CNPq (grant No. 307641/2022-8) for financial support.

\end{document}